\documentclass[aps,pra,superscriptaddress,floatfix,showpacs,10pt,twocolumn]{revtex4}
\usepackage[dvips]{graphicx}
\usepackage{amsmath}
\begin{document}
\title{Entanglement concentration for unknown atomic entangled states via entanglement swapping }
\author{Ming Yang}
\email{mingyang@ahu.edu.cn}
\author{Yan Zhao}
\author{Wei Song}
\author{Zhuo-Liang Cao}
\email{zlcao@ahu.edu.cn} \affiliation{Anhui Key Laboratory of
Information Material {\&} Devices,School of Physics {\&} Material
Science, \\Anhui University, Hefei, 230039, PRChina}
\begin{abstract}
An entanglement concentration scheme for unknown atomic
entanglement states is proposed via entanglement swapping in
cavity QED. Because the interaction used here is a large-detuned
one between two driven atoms and a quantized cavity mode, the
effects of the cavity decay and thermal field have been
eliminated. These advantages can warrant the experimental
feasibility of the current scheme.
\end{abstract}
\pacs{03.67.Hk, 03.67.Mn, 03.67.Pp} \maketitle

In quantum teleportation, an unknown quantum state will be sent
from sender to receiver via a quantum channel with the help of
classical communication. In the process, the carrier of the
unknown quantum information (state) is not transmitted. From the original discussion of quantum teleportation~\cite%
{teleportation}, a perfect quantum channel is needed if we want to
teleport the quantum state faithfully, i.e. the quantum channel
must be a maximally entangled state. But in real experiment, the
actually available quantum channels are usually non-maximally
entangled states, such as pure non-maximally entangled states and
mixed entangled states etc. To realize the faithful quantum
teleportation, we must convert the non-maximally entangled states
into the pure maximally entangled states. This process are usually
named as entanglement concentration (pure non-maximally entangled
states case)~\cite{concentration} and entanglement purification or
distillation (mixed entangled states case)~\cite{purification}.
The original contribution on entanglement purification has
discussed the general purification and concentration processes.
The main idea of the original contribution is that, from the local
operations on the ensemble of non-maximally entangled pairs (raw
entangled pairs), we will sacrifice some pairs of the raw
entangled pairs and keep the remaining ones. Then entanglement of
the remaining pairs will be enhanced. To realize the entanglement
purification process experimentally, many physical schemes have
been proposed. For mixed entangled states of polarization photons,
Pan \emph{et al} have presented their physical scheme~\cite{pan1}
and experimental scheme~\cite{pan2}, where polarization beam
splitter is the most important element. J. L. Romero \emph{et al}
proposed a physical scheme for the purification of the mixed
entangled cavity fields, where the entanglement degradation is
caused by the cavity decay. To realize the purification process,
two auxiliary atoms have been introduced~\cite{romero}. For pure
non-maximally entangled states of polarization photons, S. Bose
\emph{et al} have proposed a concentration scheme via entanglement
swapping~\cite{bose}.

In the previous contributions to entanglement purification and
concentration, there is fewer schemes discussing the purification
and concentration of entangled atomic
states~\cite{me1,me2,me3,me4}. Hitherto, the purification and
concentration of entangled photon states have been demonstrated in
experiment~\cite{pan2,kwiat,zhaozhi}, but the realization of
purification and concentration of entangled atomic states are
still under way. So it is necessary to explore the experimentally
feasible physical schemes on the purification and concentration of
entangled atomic states.

In our former contributions, by using cavity QED techniques and
linear optical elements, we realized the purification and
concentration of unknown mixed entangled atomic states and pure
non-maximally entangled atomic states
respectively~\cite{me1,me2,me3,me4}.

Entanglement swapping is another method to realize the
concentration of unknown pure non-maximally entangled
states~\cite{bose}. S.Bose \emph{et al} have discussed the photon
states case. Here, we will discuss the atomic states case in
cavity QED.

In cavity QED, the cavity decay and thermal field are two vital
obstacles to the realization of various cavity QED experiments.
Following several recent contributions on
teleportation~\cite{zheng1} , entanglement
generation~\cite{zheng2, driving} and quantum logic
gates~\cite{thermal} in cavity QED, we propose this cavity QED
scheme to realize the entanglement concentration for unknown pure
two-atom non-maximally entangled states via entanglement swapping,
where the effects of cavity decay and thermal field are all
eliminated by the large-detuned interaction between driven atoms
and the cavity mode.

Next, we will discuss the concentration process in more details.
Suppose there are three spatially separate users Alice Bob and
Cliff who share two pairs of pure non-maximally entangled atoms
$1,2$ and $3,4$~\cite{footnote}:
\begin{subequations}
\begin{equation}
|\Phi \rangle _{12}=a|e\rangle _{1}|e\rangle _{2}+b|g\rangle
_{1}|g\rangle_{2},\label{initial1}
\end{equation}
\begin{equation}
|\Phi \rangle _{34}=c|e\rangle _{3}|e\rangle _{4}+d|g\rangle
_{3}|g\rangle_{4}.\label{initial2}
\end{equation}
\end{subequations}
Here atoms $1,4$ are in the hands of users Alice and Bob
respectively, and atoms $2,3$ are all in the hand of user Cliff.
Without loss of generality, we can suppose the superposition
coefficients are all real numbers.

At Cliff's station, a single-mode cavity will be introduced. Then
Cliff will send the two atoms $2,3$ into the cavity. At the same
time, the two atoms are driven by a classical field in the cavity.
The two atoms are identical, and the energy levels used are
labelled by $e,g$ and $i$. Because the transition between $i$ and
$e$ is large dutuned from the cavity frequency, the upper level
$i$ is not affected during the interaction. The interaction
Hamiltonian of this interaction system is expressed as:
\begin{align}
H& =\omega _{0}\sum_{j=2}^{3}S_{z,j}+\omega a^{+}a
\label{hamiltonian} \nonumber\\
& +\sum_{j=2}^{3}\left[ g\left( a^{+}S_{j}^{-}+aS_{j}^{+}\right)
+\Omega \left( S_{j}^{+}e^{-i\omega_{d} t}+S_{j}^{-}e^{i\omega_{d}
t}\right) \right]
\end{align}
where $\omega _{0}$, $\omega$ and $\omega_{d}$ are atomic
transition frequency ($e\leftrightarrow g$), cavity frequency and
the frequency of driving field respectively, $a^{+}$ and $a$ are
creation and annihilation operators for the cavity mode, $g$ is
the coupling constant between atoms and cavity mode,
$S_{j}^{-}=|g\rangle _{j}\langle e|,$ $S_{j}^{+}=|e\rangle
_{j}\langle g|,$ $S_{z,j}=\frac{1}{2}\left( |e\rangle _{j}\langle
e|-|g\rangle _{j}\langle g|\right)$ are atomic operators, and
$\Omega $ is the Rabi frequency of the classical field. We
consider the case where the frequency of driving field is equal to
the atomic transition frequency, $\omega _{0}=\omega_{d}$. In the
interaction picture, the evolution operator of the system
is~\cite{zheng2}:
\begin{equation}
U\left(t\right)=e^{-iH_{0}t}e^{-iH_{eff}t}, \label{totalevolution}
\end{equation}
where $H_{0}=\sum_{j=2}^{3}\Omega\left(
S_{j}^{+}+S_{j}^{-}\right)$, $H_{eff}$ is the effective
Hamiltonian. In the large detuning $\delta \gg \frac{g}{2}$ and
strong driving field $2\Omega \gg \delta ,g$ limit, the effective
Hamiltonian for this interaction can be described as
follow~\cite{zheng2}:
\begin{align}
H_{eff}& =\lambda \left[ \frac{1}{2}\sum_{j=2}^{3}\left( |e\rangle
_{j}\langle e|+|g\rangle _{j}\langle g|\right) \right.
\label{effectivehamiltonian} \nonumber\\
& \left. +\sum_{j,k=2,j\neq k}^{3}\left(
S_{j}^{+}S_{k}^{+}+S_{j}^{+}S_{k}^{-}+\text{H.c.}\right) \right]
\end{align}
where $\lambda =\frac{g^{2}}{2\delta }$ with $\delta $ being the
detuning between atomic transition frequency $\omega _{0}$\ and
cavity frequency $\omega$.

The evolution of total system can be expressed as:
\begin{widetext}
\begin{align}
& \left( a|e\rangle _{1}|e\rangle _{2}+b|g\rangle _{1}|g\rangle _{2}\right)
\left( c|e\rangle _{3}|e\rangle _{4}+d|g\rangle _{3}|g\rangle _{4}\right)
\label{evolution} \nonumber\\
& \longrightarrow ace^{-i\lambda t}|e\rangle _{1}|e\rangle
_{4}\left[ \cos \lambda t\left( \cos \Omega t|e\rangle _{2}-i\sin
\Omega t|g\rangle _{2}\right) \times \left( \cos \Omega t|e\rangle
_{3}-i\sin \Omega
t|g\rangle _{3}\right) \right.\nonumber\\
& \left. -i\sin \lambda t\left( \cos \Omega t|g\rangle _{2}-i\sin
\Omega t|e\rangle _{2}\right) \times \left( \cos \Omega t|g\rangle
_{3}-i\sin
\Omega t|e\rangle _{3}\right) \right]\nonumber\\
& +ade^{-i\lambda t}|e\rangle _{1}|g\rangle _{4}\left[ \cos
\lambda t\left( \cos \Omega t|e\rangle _{2}-i\sin \Omega
t|g\rangle _{2}\right) \times \left( \cos \Omega t|g\rangle
_{3}-i\sin \Omega t|e\rangle _{3}\right)
\right.\nonumber\\
& \left. -i\sin \lambda t\left( \cos \Omega t|g\rangle _{2}-i\sin
\Omega t|e\rangle _{2}\right) \times \left( \cos \Omega t|e\rangle
_{3}-i\sin
\Omega t|g\rangle _{3}\right) \right]\nonumber\\
& +bce^{-i\lambda t}|g\rangle _{1}|e\rangle _{4}\left[ \cos
\lambda t\left( \cos \Omega t|g\rangle _{2}-i\sin \Omega
t|e\rangle _{2}\right) \times \left( \cos \Omega t|e\rangle
_{3}-i\sin \Omega t|g\rangle _{3}\right)
\right.\nonumber\\
& \left. -i\sin \lambda t\left( \cos \Omega t|e\rangle _{2}-i\sin
\Omega t|g\rangle _{2}\right) \times \left( \cos \Omega t|g\rangle
_{3}-i\sin
\Omega t|e\rangle _{3}\right) \right]\nonumber\\
& +bde^{-i\lambda t}|g\rangle _{1}|g\rangle _{4}\left[ \cos
\lambda t\left( \cos \Omega t|g\rangle _{2}-i\sin \Omega
t|e\rangle _{2}\right) \times \left( \cos \Omega t|g\rangle
_{3}-i\sin \Omega t|e\rangle _{3}\right)
\right.\nonumber\\
& \left. -i\sin \lambda t\left( \cos \Omega t|e\rangle _{2}-i\sin
\Omega t|g\rangle _{2}\right) \times \left( \cos \Omega t|e\rangle
_{3}-i\sin \Omega t|g\rangle _{3}\right) \right]
\end{align}
\end{widetext}
After the interaction time $t$, Cliff will measure the atoms
$2,3$. If the result is $|e\rangle _{2}|e\rangle _{3}$, the state
of atoms $1,4$ will collapse into:
\begin{align}
|\Psi \rangle _{14}& =e^{-i\lambda t}\left( e^{-i\lambda t}\cos
^{2}\Omega t+i\sin \lambda t\right) ac|e\rangle _{1}|e\rangle
_{4}\nonumber\\
&-e^{-i\lambda t}\left( e^{-i\lambda t}\sin ^{2}\Omega t+i\sin \lambda t\right) bd|g\rangle _{1}|g\rangle _{4}  \label{ee}\nonumber\\
& -e^{-2i\lambda t}\sin 2\Omega t\frac{i}{2}ad|e\rangle _{1}|g\rangle _{4}\nonumber\\
& -e^{-2i\lambda t}\sin 2\Omega
t\frac{i}{2}bc|g\rangle_{1}|e\rangle _{4}
\end{align}
From the result in Equation (\ref{ee}), we find that if we select
the atomic velocity to meet the condition $\lambda t=\frac{\pi
}{4}$, we can get a conclusive interaction time. Through
modulating the classical field, we can let the Rabi frequency
$\Omega $ satisfy the condition $\Omega t=\pi $. After the
interaction and measurements, the state of atoms $1,4$ will
become:
\begin{equation}
|\Psi \rangle _{14}=\frac{1}{\sqrt{2}}e^{-i\frac{\pi }{4}}\left(
ac|e\rangle_{1}|e\rangle _{4}-ibd|g\rangle _{1}|g\rangle
_{4}\right)  \label{result1}
\end{equation}
If the initial states of atoms $1,2$ and $3,4$ satisfy the
condition: $a=d,b=c$, the state in Equation (\ref{result1}) will
become a maximally entangled state after a rotation operation:
\begin{equation}
|\Psi \rangle _{14}=ab\frac{\left( |e\rangle _{1}|e\rangle
_{4}+|g\rangle_{1}|g\rangle _{4}\right) }{\sqrt{2}}
\label{result}
\end{equation}
The probability for obtaining this state is $P=a^{2}(1-a^{2})$.

Through analysis, we find that the measurement result $|g\rangle
_{2}|g\rangle _{3}$ also can lead to the maximally entangled state
between atom $1$ and atom $4$. But, for the other two measurement
results $|g\rangle _{2}|e\rangle _{3},|e\rangle _{2}|g\rangle
_{3}$, the atoms $1,4$ will be left in the following non-maximally
entangled state($|e\rangle _{2}|g\rangle _{3}$ case as example):
\begin{equation}
|\Psi ^{^{\prime }}\rangle
_{14}=\frac{1}{\sqrt{2}}e^{-i\frac{\pi}{4}}\left( a^{2}|e\rangle
_{1}|g\rangle _{4}-ib^{2}|g\rangle _{1}|e\rangle _{4}\right)
\label{waste}
\end{equation}
Obviously, the degree of entanglement of the state in Equation
(\ref{waste}) is smaller than the initial one. So the contribution
of the measurement results $|g\rangle _{2}|e\rangle _{3},|e\rangle
_{2}|g\rangle _{3}$ is eliminated. Totally, the successful
probability for the concentration process is
$P_{total}=2a^{2}(1-a^{2})$. Because only the measurement results
$|g\rangle _{2}|g\rangle _{3}$ and $|e\rangle _{2}|e\rangle _{3}$
are useful, it is not necessary to distinguish the two atoms $2,3$
after they flying out of the cavity.

From the above process, the two atoms $1,4$, never interacting
with each other before, are left in a pure maximally entangled
state after entanglement swapping. The entanglement swapping for
atoms has been realized by the interaction between driven atoms
and quantized cavity mode in the large detuning limit. A two-atom
maximally entangled pair has been concentrated from two unknown
atomic entangled pairs.

We have supposed the initial condition $a=d,b=c$. If there is an
error $d=a+\Delta a$, $\Delta a=ka$ with $k$ being a small
constant, the fidelity of the obtained state of atoms $1,4$ is:

\begin{equation}
F=\frac{\left[ \sqrt{1-a^{2}\left( 1+k\right) ^{2}}+\left(
1+k\right) \sqrt{1-a^{2}}\right] ^{2}}{2\left[ 1+\left(
1+k\right)^{2}-2a^{2}\left( 1+k\right) ^{2}\right] }
\label{newfidelity}
\end{equation}
relative to the state in Equation (\ref{result}). If $a=0.7,k=0.1$
the fidelity can reach $0.989$. So we can conclude that the small
error of the coefficients of the two initial states only affect
the fidelity of the result state slightly.

Next, we will discuss the experimental feasibility of the proposed
scheme. Because the interaction used here is a large-detuned one
between two driven atoms and a quantized cavity mode, the effects
of the cavity decay and thermal field have been eliminated. So we
only need consider radiative time of atoms, i.e. the whole
concentration process must be completed within the radiative time
of atoms. For the Rydberg atoms with principal quantum numbers
$50$ and $51$, the radiative time is $T_{a}=3\times 10^{-2}s$.
From the analysis in reference~\cite{zheng3}, the interaction time
is on the order $t\simeq 2\times 10^{-4}s$, which is much shorter
than the atomic radiative time $T_{a}$. So our scheme is
realizable by using cavity QED techniques.

Although it is not easy to realize the scheme experimentally, we
still can see the perspective from the researches implemented by
the group led by Haroche. The estimated interaction time is on the
order of ms, much shorter than the radiative time of the Rydberg
atoms. The required interaction time is within what can be
achieved by the experimental techniques of the group led by
Haroche. The Haroche group has realized some quantum information
processing procedures in experiment~\cite{Osnaghi, Hagley,
Rauschenbeutel1, Rauschenbeutel2}. In the cavity QED experiments,
besides the effects of the cavity decay and thermal fields, there
are still two main points that we should pay close attention to.
They are the control of the flying of the atoms and the detection
of the states of atoms. Firstly, we will discuss the control of
the flying of the atoms. From the experiment by S. Osnaghi et
al~\cite{Osnaghi}, we get that it can be realized that two atoms
at a time cross a cavity and interact with the field via virtual
photon processes. Here the two atoms have different velocities,
but they are sent at different times so that they can
simultaneously cross the cavity axis, where the two atoms can feel
the same coupling constants. At other location, the two coupling
constants are different and decided by their velocities. From the
point of average, the effect is just the same as the identical
coupling constant case. In addition, because the control of the
velocity of an atom is very difficult, ones usually control the
interaction by adjusting the atom-cavity detuning~\cite{Osnaghi}.
Secondly, we will discuss the detection of the states of atoms. In
the experiments of Haroche's group~\cite{Osnaghi, Hagley,
Rauschenbeutel1, Rauschenbeutel2}, the atomic states can be
detected by the ionization detectors with high efficiency. If the
detector fails to detect the atom, the experiment restarts. This
decreases the probability of success, but does not affect the
fidelity of the obtained entangled states. Therefore, the atomic
detection is not a problem in our scheme.

In the current scheme, the two atoms must be sent through the
cavity simultaneously. But in real experiment, there exist some
errors. Further calculation on the error suggests that this error
only affects the fidelity of the result state slightly.

In conclusion, we have presented an entanglement concentration
scheme for unknown pure two-atom non-maximally entangled states
using entanglement swapping. Because, in the scheme, we use the
interaction between driven atoms and quantized cavity mode in the
large detuning limit, the scheme embeds the advantage that it is
insensitive to the cavity decay and thermal state; the initial
state of the pure non-maximally entangled atoms is unknown, i.e.
the concentration process can deal with the unknown input states
successfully; In addition, there is no need for us to distinguish
the two atoms $2,3$ when they fly out of the cavity area after
interaction.

\begin{acknowledgements}
This work is supported by the Natural Science Foundation of the
Education Department of Anhui Province under Grant No: 2004kj005zd
and Anhui Provincial Natural Science Foundation under Grant No:
03042401 and the Talent Foundation of Anhui University.
\end{acknowledgements}


\begin{thebibliography}{99}
\bibitem{teleportation} C. H. Bennett, G. Brassard, C. Cr\'{e}peau, R.
Jozsa, A. Peres, \& W. K.Wootter, Phys. Rev. Lett. 70, 1895
(1993).

\bibitem{concentration} C. H. Bennett, H. J. Bernstein, S. Popescu \& B.
Schumacher, Phys. Rev. A 53 2046 (1996).

\bibitem{purification} C. H. Bennett, G. Brassard, S. Popescu, B.
Schumacher, J. A. Smoin, \& W. K. Wootters, Phys. Rev. Lett. 76, 722 (1996).

\bibitem{pan1} J. W. Pan, C. Simon, \v{C}. Brukner \& A. Zeilinger, Nature
410, 1067 (2001).

\bibitem{pan2} J. W. Pan, S. Gasparoni, R. Ursin, G. Weihs \& A. Zeilinger,
Nature 423, 417 (2003).

\bibitem{romero} J. L. Romero, L. Roa, J. C. Retamal, \& C.Saavedra, Phys.
Rev. A 65, 052319 (2002).

\bibitem{bose} S. Bose, V. Vedral, \& P. L. Knight, Phys. Rev. A 60, 194
(1999).

\bibitem{me1} M. Yang, \& W. Song, Z. L. Cao, Phys. Rev. A 71, 012308 (2005).

\bibitem{me2} M. Yang, \& W. Song, Z. L. Cao, Physica. A, 341 (2004) 251-261. M. Yang, \& Z. L. Cao, Physica. A, 337/1-2, (2004) 141-148.

\bibitem{me3} Z. L. Cao, \& M. Yang, J. Phys. B 36, 4245 (2003).

\bibitem{me4} Z. L. Cao, \& M. Yang, G. C. Guo, Phys Lett A 308 349 (2003).

\bibitem{kwiat} P. G. Kwiat, S. Barraza-Lopez, A. Stefanov \& N. Gisin,
Nature, Vol. 409, 1014 (2001).

\bibitem{zhaozhi} Z. Zhao, T. Yang, Y.-A. Chen, A.-N. Zhang, and J.-W. Pan,
Phys. Rev. Lett. 90, 207901 (2003).

\bibitem{zheng1} S.-B. Zheng, Phys. Rev. A 69, 064302 (2004).

\bibitem{zheng2} S.-B. Zheng, Phys. Rev. A 68, 035801 (2003).

\bibitem{driving}E. Solano, G. S. Agarwal, and H. Walther, Phys. Rev. Lett. 90, 027903(2003).

\bibitem{thermal}S.-B. Zheng, Phys. Rev. A 66, 060303 (2002).

\bibitem{footnote} This kind of pur\ non-maximally entangled atomic states may be
generated from a non-perfect generation process.

\bibitem{zheng3} S-B. Zheng and G-C. Guo, Phys. Rev. Lett. 85, 2392 (2000).

\bibitem{Osnaghi}S. Osnaghi , P. Bertet, A. Auffeves, P. Maioli, M. Brune, J. M. Raimond, and S.
Haroche, Phys. Rev. Lett. 87, 037902 (2001)

\bibitem{Hagley}E. Hagley, X. Maitre, G. Nogues, C. Wunderlich, M. Brune, J. M. Raimond, S.
Haroche, Phys. Rev. Lett. 79, 1 (1997).


\bibitem{Rauschenbeutel1}A. Rauschenbeutel , G. Nogues, S. Osnaghi, P. Bertet, M. Brune, J. M. Raimond,
and S. Haroche,Phys. Rev. Lett. 83, 5166 (1999).


\bibitem{Rauschenbeutel2}A. Rauschenbeutel, G. Nogues, S. Osnaghi, P. Bertet, M. Brune, J. M. Raimond,
Serge Haroche, Science 288, 2024 (2000).

\end{thebibliography}
\end{document}